\documentstyle[prd,aps,epsfig,eqsecnum,amssymb,amsmath,floats]{revtex}
\begin{document}
\title{Neutrino background spectra from primordial black hole evaporations:
dependence on the initial mass spectrum.}
\author{
E. V. Bugaev and K. V. Konishchev
}
\address{
Institute for Nuclear Research, Russian Academy of Sciences, 
Moscow 117312, Russia
}
\date{
\today
}
\maketitle
\begin{abstract}
Comparative analysis of different types of initial mass spectra of
primordial black holes (PBHs) is carried out. It is assumed that primordial density 
fluctuations have power law spectrum and the spectral index is constant throughout
all scales. It is shown that densities of background radiations ($\nu$, $\gamma$)
from PBH evaporations strongly depend on the type of gravitational collapse
and on the way of taking into account the spread of horizon masses at which 
PBHs can form. Constraints on spectral index values based on PBH evaporation 
process and on atmospheric and solar neutrino data are obtained.
\end{abstract}

\section{Initial  mass spectra of PBHs}

The initial mass function of primordial black holes (or, it is better
to say, the initial mass spectrum) is determined, first of all, by the
peculiarities of a gravitational collapse near the threshold of a black
hole formation. According to analytic calculations of seventies \cite{1,2}
a critical size of the density contrast needed for the PBH formation, $\delta_c$,
is about $1/3$. Besides, it was argued that all PBHs have mass roughly
equal to the horizon mass at a moment of the formation, independently
of the perturbation size. However, it was shown recently
that near the threshold of black hole formation the gravitational collapse
behaves as a critical phenomenon \cite{3}. In this case the initial mass function
will be quite different from the analogous function in the analytic calculations \cite{1,2}.
The main feature is that the PBH mass formed depends on the size of the
fluctuation \cite{3},
\begin{equation}
\label{1}
M_{BH}=k M_h (\delta-\delta_c)^{\gamma_k}.
\end{equation}
In this formula $M_h$ is the horizon mass at the time the fluctuation
enters the horizon; $\delta_c$, $k$, $\gamma$ are parameters of a concrete
model of the critical collapse \cite{3,4}. It is seen from Eq.(\ref{1})
that PBH mass may be arbitrarily small independently of the value of $M_h$.
Besides, the value of the critical overdensity, $\delta_c$, in such models
is typically $\sim 0.7$, i.e., about a factor of 2 larger than the value 
found analytically \cite{2}.

The second important ingredient of a PBH initial mass function calculation
is a taking into account the spread of horizon masses at which PBHs are formed. 
This problem exists independently of a nature of gravitational collapse.
It is assumed very often that the majority of the black hole formation occurs
at the shortest possible scale. In particular, the authors of ref.\cite{3}
determined the PBH initial mass function under the assumption
that all PBHs form at the same horizon mass. The accuracy of such approximation
was studied in the work \cite{5} using the excursion set formalism, 
and it was found that it is good enough, at least in the case of power-law
density perturbation spectra with the spectral index close to $1.2$ (i.e.,
slightly "blue" spectra, with the value of $n$ satisfying observational constraints).

Recently, in the works of Kim \cite{6} and present authors \cite{7}
the expression for the PBH mass spectrum (for the critical collapse case) 
was obtained which takes into account
the accumulation of PBHs  formed at all epochs after the beginning of
the ordinary radiation dominated era. The derivation is based on the Press-
Schechter
formalism \cite{8}. Resulting formula for the PBH mass spectrum,
taking into account the spread of horizon masses, depends on
the horizon crossing amplitude $\sigma_H$ which is a function of horizon mass $M_h$.
In the present paper we assume a power law  of primordial density perturbations
for all scales, and in this case one has ($\Omega_{\Lambda} = 0$)
\begin{equation}
\label{2}
\sigma_H(M_h)=C_n\delta_H(M_h),
\end{equation}
\begin{equation}
\label{3}
\delta_H(M_h)\cong 2\cdot 10^{-5} \left(\frac{M_{eq}}{M_{h\!0}}\right)^{\frac{1-n}{6}}
\left(\frac{M_h}{M_{eq}}\right)^{\frac{1-n}{4}}.
\end{equation}

In these formulas $\delta_H(M_{h})$ is the non-smoothing horizon crossing amplitude
(which is time-independent), $M_{h\!0}$ is the present horizon mass, $M_{eq}$ is the horizon
mass at the moment of matter-radiation equality, $C_n$ is the numerical coefficient
arising as a result of the smoothing procedure (the top-hat form of the window-function
was used in all numerical calculations). The amplitude is normalized on COBE data 
(see \cite{7} for detailes and references).

The expression for the PBH mass spectrum is \cite{7}
\begin{equation}
\label{4}
n_{BH}(M_{BH})=\frac{n+3}{4}\sqrt{\frac{2}{\pi}}\rho_i M_i^{1/2} M_{BH}^{-5/2}\int
\limits_{\delta_c}^{\delta_{max}}\frac{1}{\sigma_H} \left|\frac{\delta^2}{\sigma_H^2}-1\right|
e^{-\frac{\delta^2}{2\sigma_H^2}}\xi^{3/2} d\delta,
\end{equation}
$$
\xi\equiv k(\delta-\delta_c)^{\gamma_k}.
$$
Here, $\rho_i$ amd $M_i$ are the background density and the horizon mass at $t_i$
($t_i$ is the moment of a beginning of growth of the density fluctuations). The value of PBH
mass $M_{BH}$ is connected with $M_h$ by the relation (\ref{1}). The upper limit of 
integration in Eq.(\ref{4}) is determined by the expression
\begin{equation}
\label{5}
\delta_{max}=min\left[\left(\frac{M_{BH}}{k M_{i}}\right)^{1/{\gamma_k}}\!\!\!\!+\delta_c,\,\,\,1\right].
\end{equation}
Note, that in ref.\cite{7} the upper limit of integration in the expression for $n_{BH}(M_{BH})$ 
was erroneously put equal to 1.

All numerical calculations were carried out with the following values of parameters \cite{3}:
\begin{equation}
\label{6a}
k=3\;\;\;,\;\;\; \gamma_k=0.36\;\;\;,\;\;\; \delta_c=0.7.
\end{equation}

The known formula \cite{9} for the PBH mass spectrum in the  standard collapse case can be obtained from Eq.(\ref{4}) by
using the substitutions \cite{7}
\begin{equation}
\label{6b}
\gamma_k \to 0 \;\;\;,\;\;\; k\to \gamma^{1/2}\;\;\;,\;\;\; \delta_c \to \gamma
\end{equation}
and the approximate relation 
\begin{equation}
\label{6c}
\int^1_{\gamma} d\delta ' \left(\frac{\delta '{}^2}{\sigma_H^2} -1 \right) e^{-\frac{\delta '{}^2}{2\sigma_H^2}}
\approx \gamma e^{-\frac{\gamma^2}{2\sigma_H^2}}\;\;.
\end{equation}
Substituting Eqs.(\ref{6b}) and (\ref{6c}) in Eq.(\ref{4}) one  obtains
\begin{equation}
\label{6d}
n_{BH}(M_{BH})=\frac{n+3}{4}\sqrt{\frac{2}{\pi}}\gamma^{7/4}\rho_iM_i^{1/2}
M_{BH}^{-5/2}\sigma_H^{-1}\exp\left(-\frac{\gamma^2}{2\sigma_H^2}\right)\;.
\end{equation}
In this case the relation between $M_{BH}$ and $M_{h}$ is independent  on $\delta$:
\begin{equation}
\label{6e}
M_{BH}=\gamma^{1/2}M_h.
\end{equation}

We  carried out in this work the calculations of the neutrino background spectra from PBH
evaporations using the mass spectrum given by Eq.(\ref{4}) and, in parallel, the mass spectrum
from ref.\cite{3},
\begin{equation}
\label{6}
n_{BH}(M_{BH})=\frac{\rho_i\left(\frac{M_{BH}}{k M_i}\right)^{1/{\gamma_k}}}{\sqrt{2\pi}
\sigma_H M_{BH} M_i \gamma_k} e^{-\frac{\left[\left(\frac{M_{BH}}{k M_i}\right)^{1/{\gamma_k}}
+\delta_c\right]^2}{2\sigma_H^2}}.
\end{equation}
In this expression the amplitude $\sigma_H$ is determined by the same formula (\ref{2}),
as above, but with the constant value of $M_h$, $M_h=M_i$, in accordance with the assumption
of authors of ref.\cite{3} that all PBHs form at the smallest scale.  

Our approach has 
two free parameters: the spectral index $n$, giving the perturbation amplitude (normalized
on COBE data on large scales) and $t_i$, the moment of time just after reheating, from
which the process of PBHs formation started. The value of $t_i$ is connected with a value of the
reheating temperature,
\begin{equation}
\label{7}
t_i=0.301 g_*^{-1/2}\frac{M_{pl}}{T_{RH}^2}
\end{equation}

($g_*\sim 100$ is the number of degrees of freedom in the early universe).

Typical results of PBH mass spectrum calculations are shown on Fig.1.
We use, for convenience, the following abbreviations: KL, NJ and BK signify, correspondingly,
the PBH mass spectra calculated by formulae of Kim and Lee \cite{9} (Eq.(\ref{6d})), Niemeyer and Jedamzik \cite{3} (Eq.(\ref{6}))
and Bugaev and Konishchev \cite{7} (Eq.(\ref{4})).
One can see from Fig.1 that at large values of the spectral index  there is a clear difference
in a behavior of two mass spectra, NJ and BK, based both on a picture of the critical collapse:
BK spectrum is less steep at $M_{BH}>M_{i}$. Of  course, this difference  is not so spectacular in cases when
the spectral index $n$ is more close to 1 (it is proved by another  method in ref.\cite{5}).

The common feature of NJ and BK spectra is the maximum at $M_{BH}\cong M_i$. Besides, both spectra have same slope
in the region $M_{BH}<M_i$.

\section{Neutrino background spectra from PBHs evaporations}

Evolution of a PBH mass spectrum due to the evaporation leads to the
approximate expression for this spectrum at any moment of time:
\begin{equation}
\label{8}
n_{BH}(m,t)=\frac{m^2}{(3\alpha t + m^3)^{2/3}} n_{BH} \left((3\alpha t + m^3)^{1/3}\right),
\end{equation}
where $\alpha$ accounts for the degrees of freedom of evaporated particles and, strictly 
speaking, is a function of a running value of the PBH mass $m$. In all our numerical
calculations we use the approximation
\begin{equation}
\label{9}
\alpha=const=\alpha (M_{BH}^{max}),
\end{equation}
where $M_{BH}^{max}$ is the value of $M_{BH}$ in the initial mass spectrum
corresponding to the maximum of this spectrum. Special study shows that errors
connected with such approximation are rather small.

The expression for the spectrum of the background  radiation is \cite{7}
\begin{eqnarray}
\label{10}
S(E)=\frac{c}{4\pi}\int dt \frac{a_0}{a}\left(\frac{a_i}{a_0}\right)^{3}
\int dm \frac{m^2}{(3\alpha t + m^3)^{2/3}} n_{BH} \left[(3\alpha t + m^3)^{1/3}\right]
\cdot f(E\cdot (1+z),m)e^{-\tau(E,z)}\nonumber\\
\\
\equiv \int F(E,z)d \log_{10} (z+1).\nonumber
\end{eqnarray}

In this formula $f(E,m)$ is a total instantaneous spectrum of the background radiation
(neutrinos or photons) from a black hole evaporation. It includes the pure Hawking 
term and contributions from fragmentations of evaporated quarks and from decays of
pions and muons (see \cite{7,7a} for details). The exponential factor in Eq.(\ref{10})
takes into account an absorption of the radiation during its propagation in
space. The processes of the neutrino absorption are considered, in a given context,
in ref.\cite{7a}. 

In last line of Eq.(\ref{10}) we  changed the variable $t$ on $z$ using, for simplicity,
the flat model with $\Omega_{\Lambda}=0$
 for which
\begin{eqnarray}
\label{11}
\frac{dt}{dz}=-\frac{1}{H_0 (1+z)}\left(\Omega_m (z+1)^{3}+\Omega_r (z+1)^{4}\right)
^{-1/2},\nonumber\\
\\
\Omega_r = (2.4\cdot 10^{4} h^2)^{-1} \;\;\;,\;\;\; h=0.67.\nonumber
\end{eqnarray}

Several examples of $z$-distributions (the integrands of the integral over $z$ in Eq.(\ref{10}))
are shown on Fig.2. Again one can see the rather strong difference of two cases: the 
existence of a tail of heavy masses in BK spectrum leads to a relative enhancement
of low $z$-contributions. The effect becomes more distinct with a rise of the reheating
temperature.
 
The sharp cut-off of all $z$-distributions near $z\sim 10^7$ is entirely due to the neutrino
absorption 
\cite{7a}. The shrinkage of $z$-distributions at large $T_{RH}$ in NJ-case is due to the absence
of large masses in the spectrum (PBHs of small masses evaporated earlier, and their radiation today is more redshifted).

Some results of  calculations of the neutrino background spectra are shown on Figs.3-5.
The functional form of these spectra is qualitatively the same as in the standard collapse case \cite{7a}:
$E^{-3}$ law at large neutrino energies and a flat part at low energies with the crossover energy
depending on $T_{RH}$. The only new feature is the sharp steepening of the spectra at high values
of $T_{RH}$ in NJ case (Fig.4) connected with the corresponding shrinkage of  the $z$-range.

Fig.5 shows the sensitivity of background neutrino intensities to a chosen value of the spectral index~$n$,
at a fixed $T_{RH}$. One can see from this figure that the neutrino background flux from PBH evaporations is 
comparable with the atmospheric neutrino flux at $\sim 100\text{ MeV}$ if $n\sim 1.29$ (at $T_{RH}\sim 10^{9}\text{ GeV}$).
Such large values of the spectral index are, probably, excluded by the recent large-scale experiments.

Next figure shows the constraints on the spectral index following from  data of 
neutrino experiments, as a function of the reheating temperature. As in our previous work,
we used for a description of these constraints the data of the Kamiokande atmospheric neutrino experiment \cite{12}
and the experiment on a search of an antineutrino flux from the Sun \cite{13}
(see \cite{8} for details). 

For completeness, we show on the same figure the constraints following
from extragalactic diffuse gamma ray data; part of them (NJ case) is in qualitative
agreement with the results of ref.\cite{13a}.  It is seen that the constraints
are much more weak in NJ case (at $T_{RH}\agt 10^{10}\text{ GeV}$). It is again
the result of an absence of the tail of large masses in NJ spectrum, leading
to a shrinkage of the $z$-distriburions and to a steepening of the background neutrino
spectra.

\section{Conclusions}

1. The initial concentration of PBH's is larger in the case of the standard picture of
the gravitational collapse (as compared with the critical collapse case). It is due to
relatively small value of the critical overdensity ($\delta_c^{st}=\gamma=1/3$). In the 
region near the maximum of PBH mass spectra the ratio of intensities is given by the 
approximate relation
\begin{equation}
\label{12}
\frac{n_{BH}^{(KL)}}{n_{BH}^{(BK)}}\sim e^{\frac{\delta_c^2-\gamma^2}{2\sigma_H^2}}\;.
\end{equation}
Correspondingly, densities of the neutrino and gamma backgrounds produced by PBH's evaporations
are smaller and spectral index constraints are weaker in the critical collapse case.

2. The summation over all epochs of PBH's production (i.e., the transition from NJ to BK case)
leads to an appearing of the tail of large PBH masses in the PBH mass spectrum. This tail is 
especially sufficient if the spectral index $n$ is relatively large.

3. In scenarios with high reheating temperatures, $T_{RH}\sim 10^9-10^{10} \text{GeV}$, the summation
over all epochs leads to a strong enhancement of a high energy part of the background neutrino spectra.

4. Spectral index constraints following from the comparison of neutrino background predictions with 
existing data of neutrino experiments are rather weak if the purely power law of primordial density 
fluctuations is assumed for all scales (and, correspondingly, the normalization on COBE data is used).
However, it does not mean that neutrino background intensities from PBH's evaporations cannot be
noticeable. If, for example, the spectrum of primordial density fluctuations is a combination of two simple power
law spectra \cite{6}, i.e.,
\begin{eqnarray}
\label{13}
\sigma_H(M_h)\sim M_h^{\frac{1-n_s}{4}}\;,\;\;\;\;\;M_h<M_{h_c}<M_{eq}\;;\nonumber\\
\\
\sigma_H(M_h)\sim M_h^{\frac{1-n_l}{4}}\;,\;\;\;\;\;M_{eq}>M_h>M_{h_c},\nonumber
\end{eqnarray} 
then the small value of $n_l\sim 1$ ($n_l-1\approx 0$) which follows, in particular,
from COBE data, does not contradict with the possibility that $n_s-1$ is large and,
correspondingly, $\sigma_H$ at small scales is also large. If $M_{h_c}$ and $n_l$
are known from somewhere, one can easily obtain from the neutrino data the constraints on the 
small scale spectral index $n_s$, in the same manner as it is done above for the all 
scale spectral index $n$. Several cosmological models of such kind,  which predict 
a large density fluctuation amplitude just on small scales (while
the amplitude on large scales is constrained by the small CMBR anisotropy) 
appeared recently (see, e.g., \cite{14,15}).

\begin{figure}[!t]
\epsfig{file=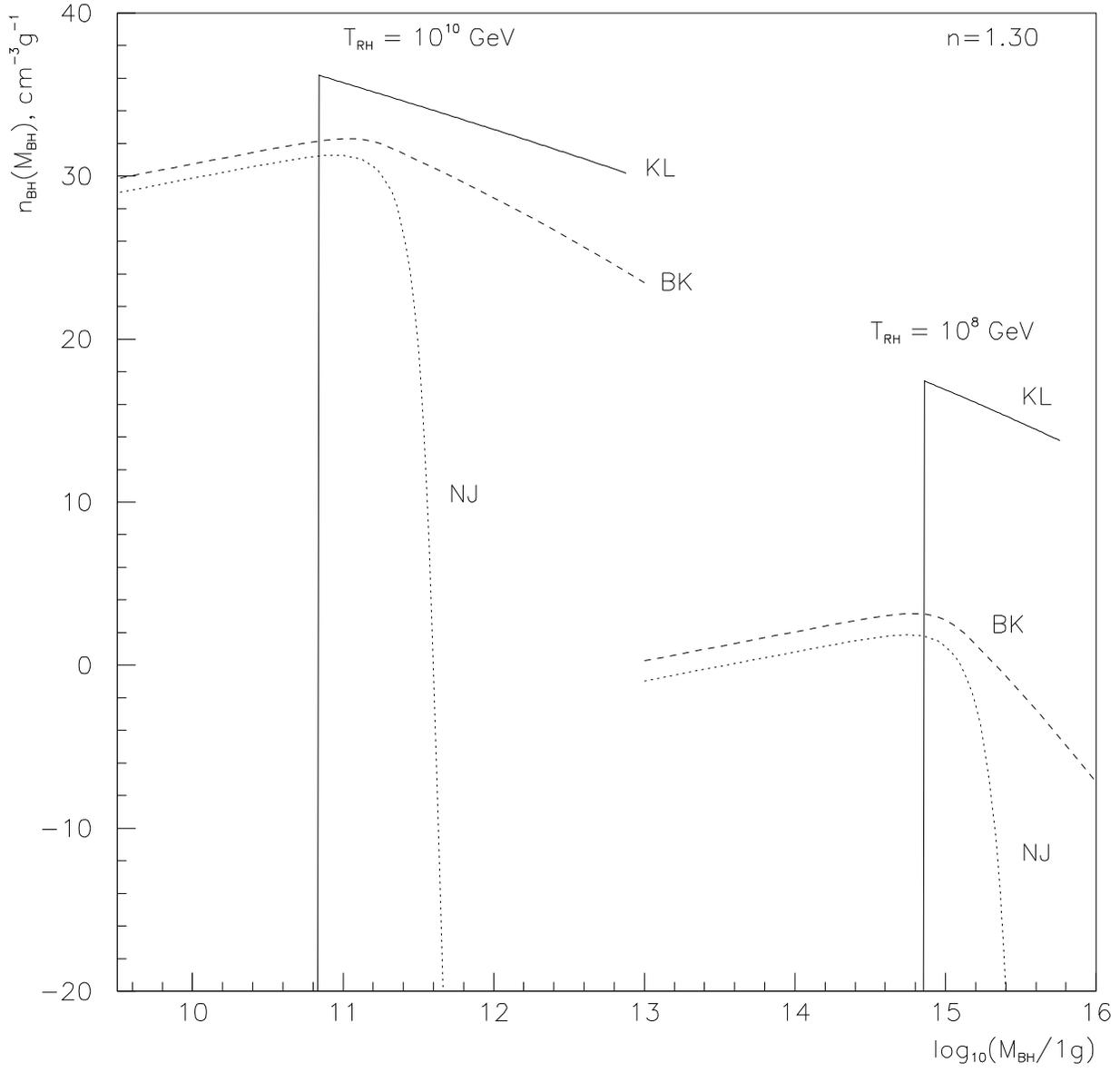,width=\columnwidth}
\caption{Examples of PBH mass spectra for two values of $T_{RH}$. Solid lines: KL spectra [10], dotted lines:
NJ spectra [3], dashed lines: BK spectra [7]; $n=1.30$ in all cases.}
\label{fig:fig1}
\end{figure}

\begin{figure}[!t]
\epsfig{file=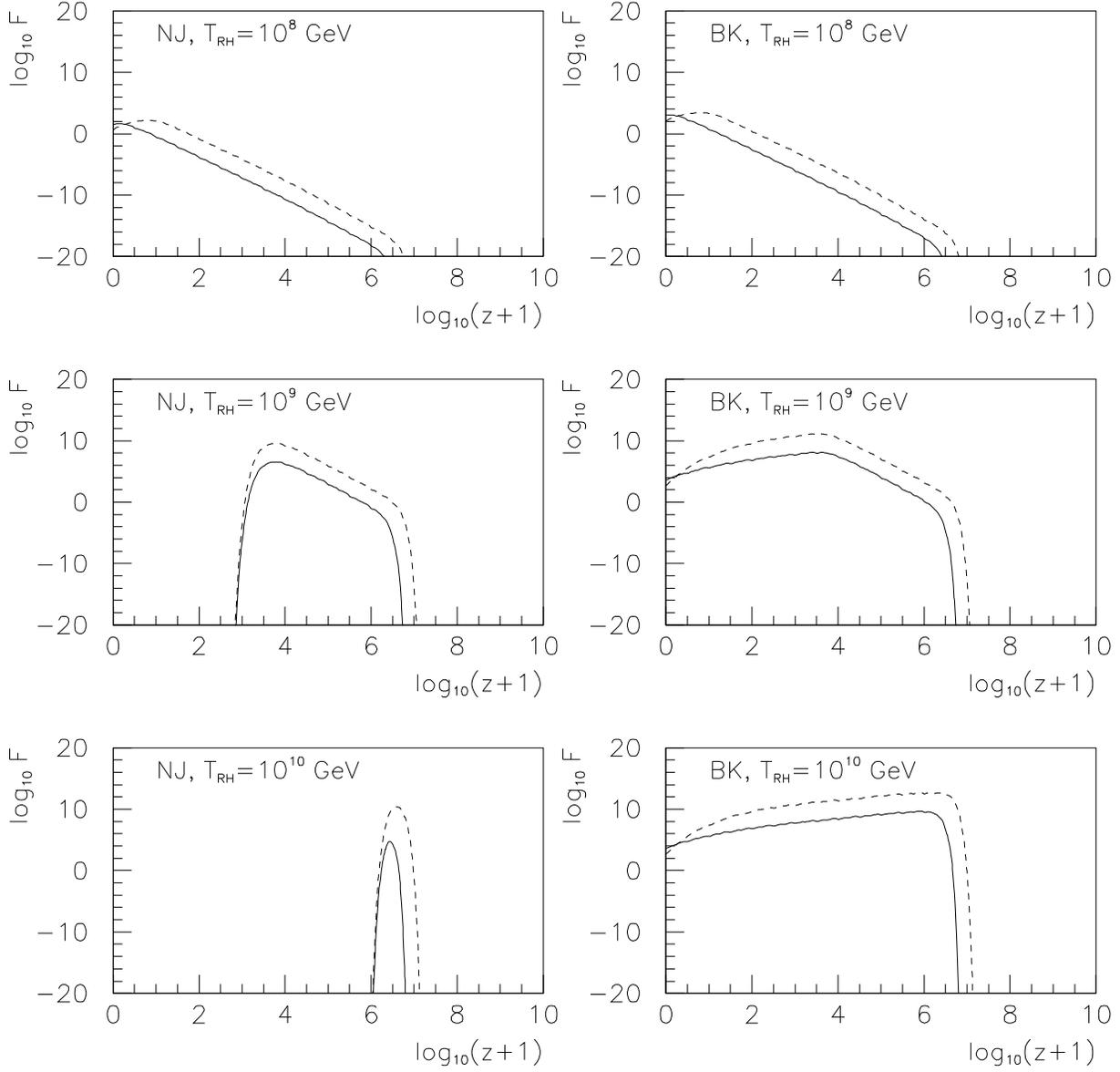,width=\columnwidth}
\caption{Red-shift distributions (integrands of the expression (\ref{10}) for the background spectrum) for neutrino energy
$E=10^{-1}\text{ GeV}$ (solid lines) and $E=10^{-2}\text{ GeV}$ (dashed lines), for PBH mass spectrum (left column) and
BK mass spectrum (right column), for three values of $T_{RH}$; $n=1.30$ in all cases. The dimension of $F(z)$ is ($s^{-1} cm^{-2} sr^{-1} GeV^{-1}$).}
\label{fig:fig2}
\end{figure}

\begin{figure}[!t]
\epsfig{file=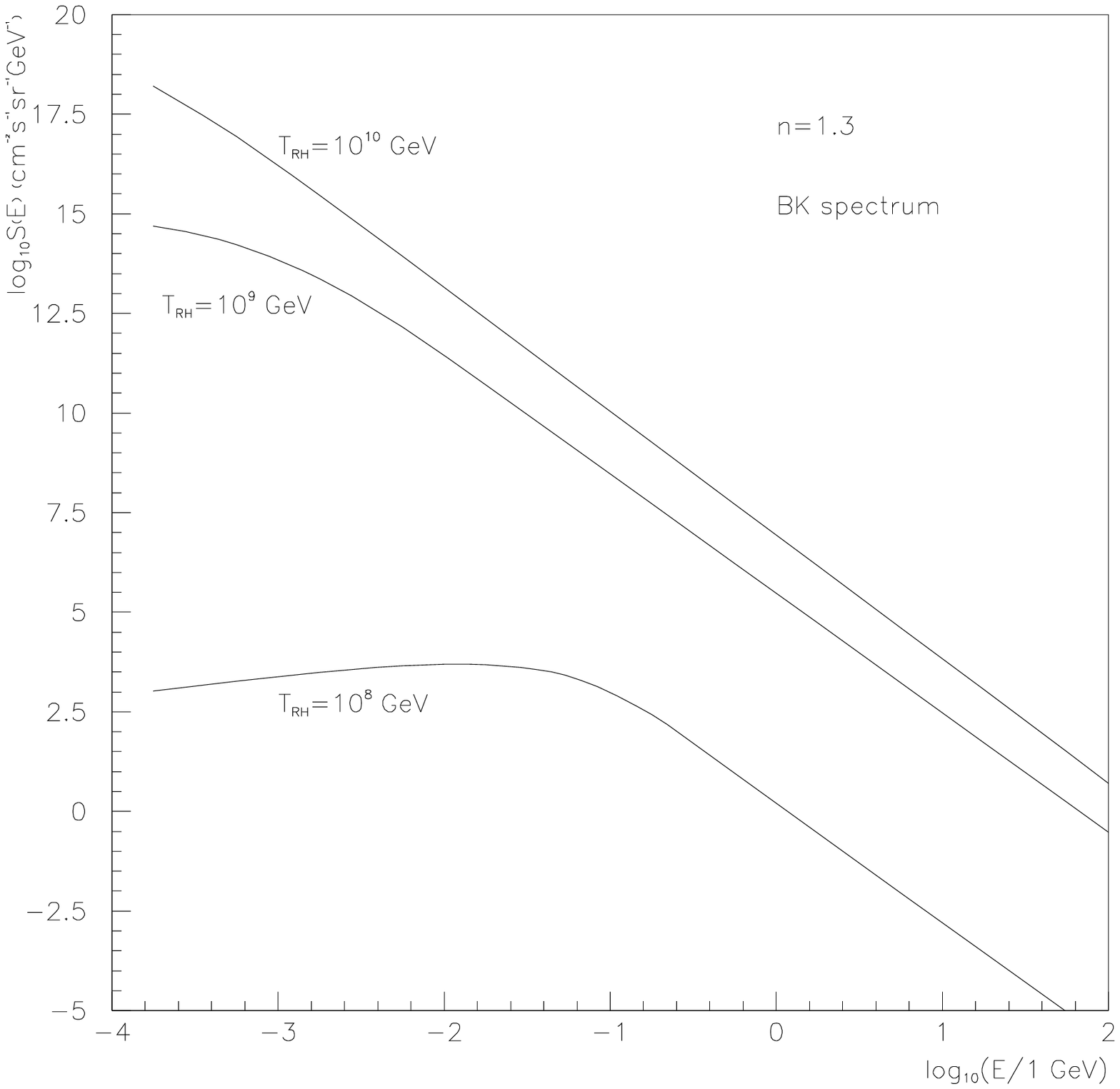,width=\columnwidth}
\caption{Neutrino background spectra for three values of $T_{RH}$, for BK mass spectrum; $n=1.30$.}
\label{fig:fig3}
\end{figure}

\begin{figure}[!t]
\epsfig{file=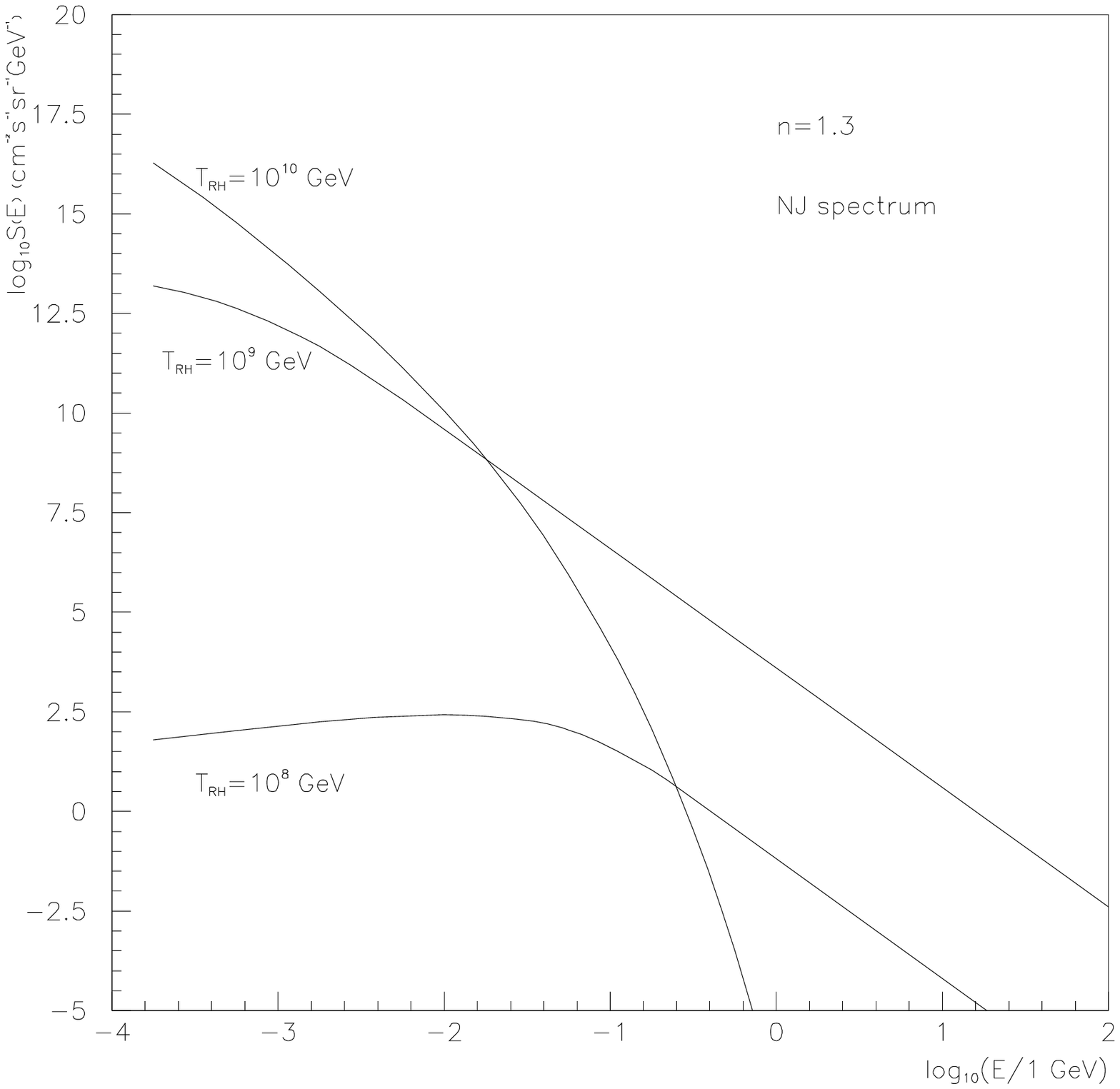,width=\columnwidth}
\caption{Neutrino background spectra for three values of $T_{RH}$, for NJ mass spectrum; $n=1.30$.}
\label{fig:fig4}
\end{figure}

\begin{figure}[!t]
\epsfig{file=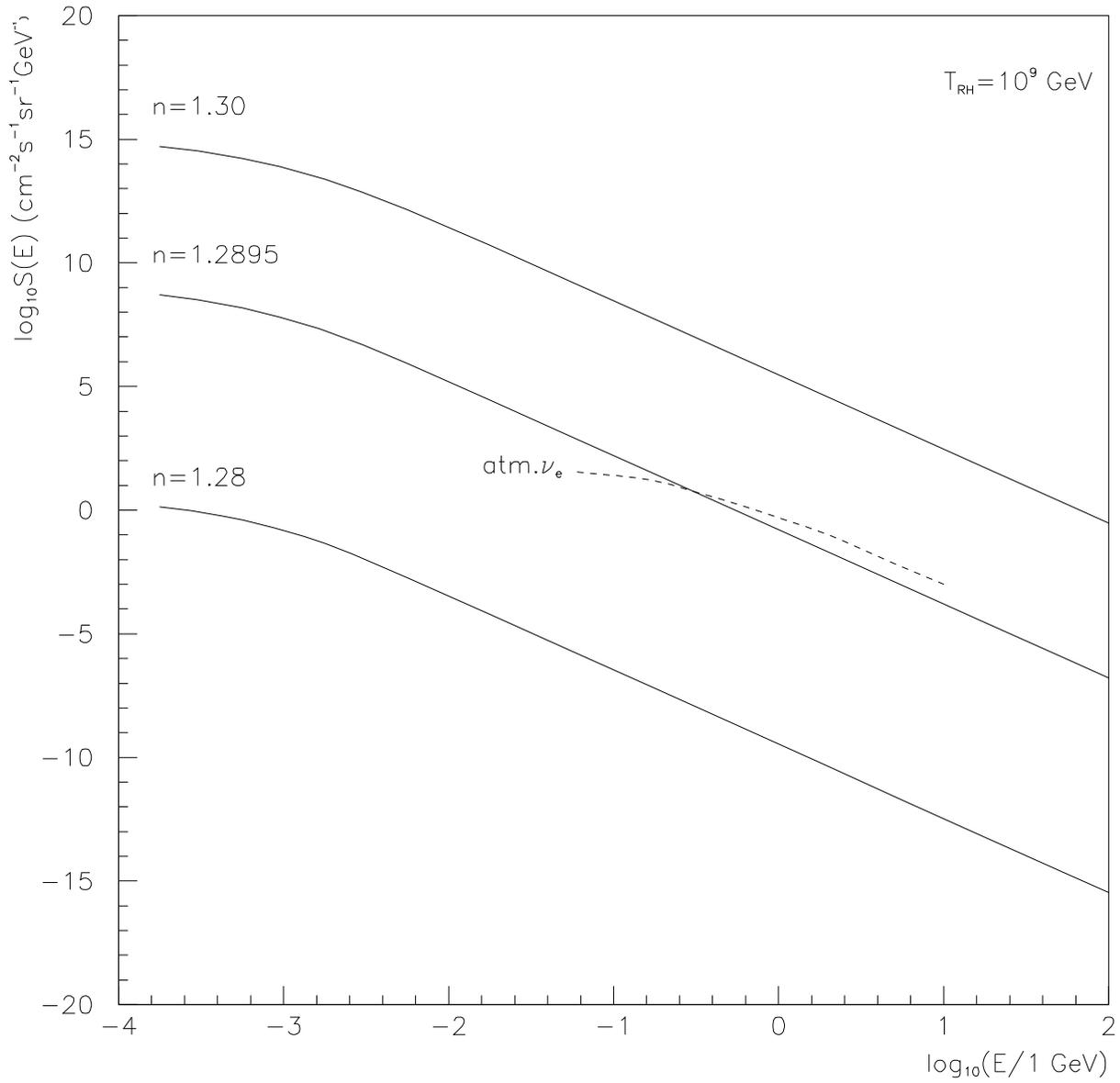,width=\columnwidth}
\caption{Neutrino background spectra for three values of the parameter $n$, for BK mass spectrum. Dashed line
represents the theoretical atmospheric neutrino spectrum for Kamiokande site (averaged over all directions) [11].}
\label{fig:fig5}
\end{figure}

\begin{figure}[!t]
\epsfig{file=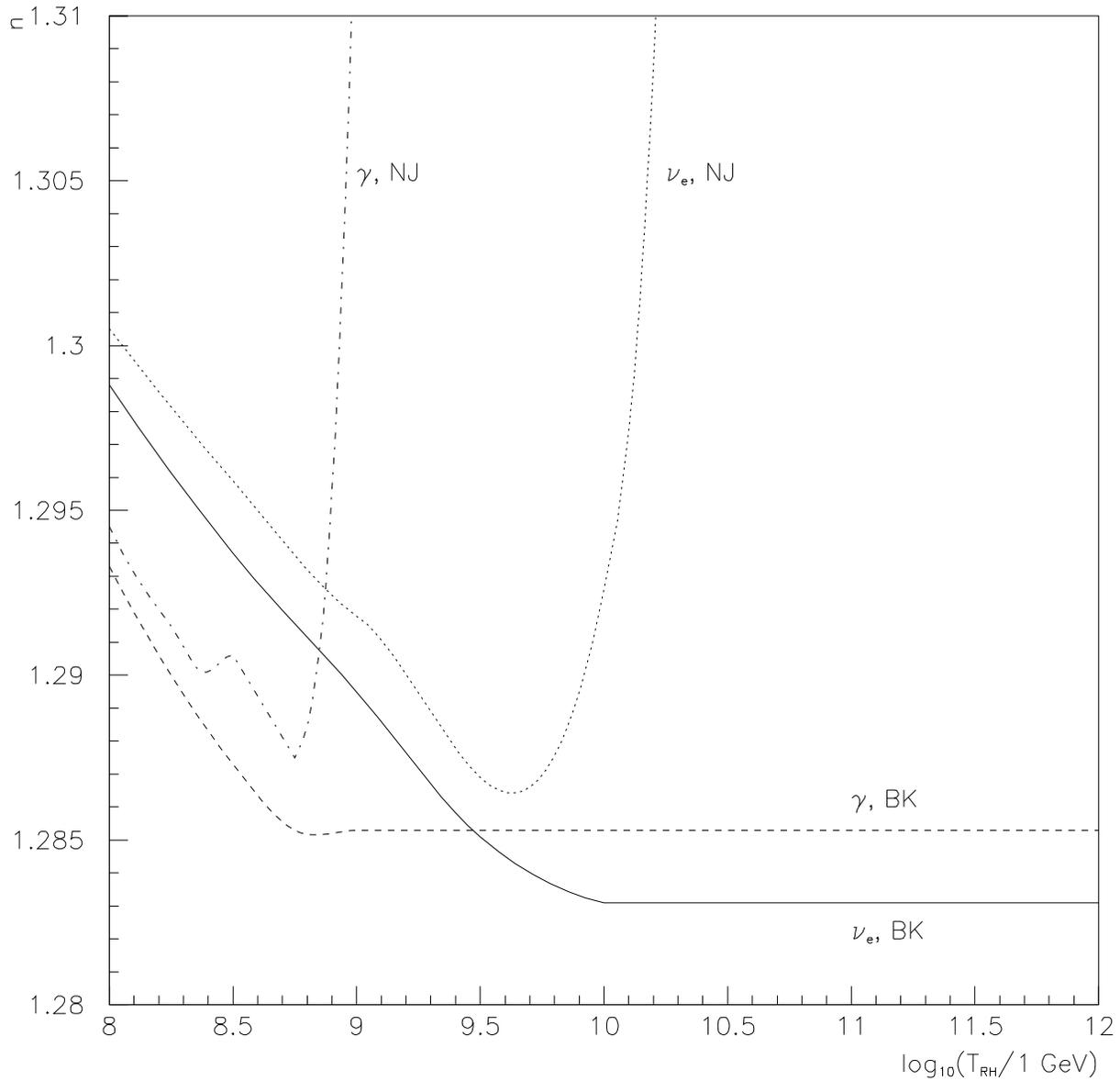,width=\columnwidth}
\caption{Constraints on the spectral index $n$ as a function of the reheating temperature $T_{RH}$. Solid and dotted lines:
BK and NJ mass spectra, atmospheric $\nu_e$ and solar $\tilde \nu_e$ experiments. Dashed and dot-dashed lines: BK and NJ mass
spectra, extragalactic diffuse gamma-ray background data.}
\label{fig:fig6}
\end{figure}

\end{document}